\documentclass[twocolumn]{aastex631}

\usepackage{gensymb}
\usepackage{soul}

\graphicspath{{./}}

\begin{document}

\title{The TESS Faint Star Search: 1,617 TOIs from the TESS Primary Mission}

\correspondingauthor{Michelle Kunimoto}
\email{mkuni@mit.edu}

\author[0000-0001-9269-8060]{Michelle Kunimoto}
\affiliation{Department of Physics and Kavli Institute for Astrophysics and Space Research, Massachusetts Institute of Technology, Cambridge, MA 02139, USA}

\author[0000-0002-6939-9211]{Tansu Daylan}
\affiliation{Department of Physics and Kavli Institute for Astrophysics and Space Research, Massachusetts Institute of Technology, Cambridge, MA 02139, USA}
\affiliation{Department of Astrophysical Sciences, Princeton University, 4 Ivy Lane, Princeton, NJ 08544}

\author[0000-0002-5169-9427]{Natalia Guerrero}
\affiliation{Department of Astronomy, University of Florida, Gainesville, FL 32611}

\author[0000-0003-0241-2757]{William~Fong}
\affiliation{Department of Physics and Kavli Institute for Astrophysics and Space Research, Massachusetts Institute of Technology, Cambridge, MA 02139, USA}

\author[0000-0003-0081-1797]{Steve Bryson}
\affiliation{NASA Ames Research Center, Moffett Field, CA 94035, USA}

\author[0000-0003-2058-6662]{George~R.~Ricker}
\affiliation{Department of Physics and Kavli Institute for Astrophysics and Space Research, Massachusetts Institute of Technology, Cambridge, MA 02139, USA}

\author[0000-0002-9113-7162]{Michael~Fausnaugh}
\affiliation{Department of Physics and Kavli Institute for Astrophysics and Space Research, Massachusetts Institute of Technology, Cambridge, MA 02139, USA}

\author[0000-0003-0918-7484]{Chelsea~ X.~Huang}
\affiliation{Department of Physics and Kavli Institute for Astrophysics and Space Research, Massachusetts Institute of Technology, Cambridge, MA 02139, USA}
\affiliation{Juan Carlos Torres Fellow}

\author[0000-0001-5401-8079]{Lizhou~Sha}
\affiliation{Department of Physics and Kavli Institute for Astrophysics and Space Research, Massachusetts Institute of Technology, Cambridge, MA 02139, USA}

\author[0000-0002-1836-3120]{Avi Shporer}
\affiliation{Department of Physics and Kavli Institute for Astrophysics and Space Research, Massachusetts Institute of Technology, Cambridge, MA 02139, USA}

\author[0000-0001-7246-5438]{Andrew~Vanderburg}
\affiliation{Department of Physics and Kavli Institute for Astrophysics and Space Research, Massachusetts Institute of Technology, Cambridge, MA 02139, USA}

\author[0000-0001-6763-6562]{Roland K.\ Vanderspek}
\affiliation{Department of Physics and Kavli Institute for Astrophysics and Space Research, Massachusetts Institute of Technology, Cambridge, MA 02139, USA}

\author[0000-0003-1667-5427]{Liang~Yu}
\affiliation{Department of Physics and Kavli Institute for Astrophysics and Space Research, Massachusetts Institute of Technology, Cambridge, MA 02139, USA}

\begin{abstract}

We present the detection of 1,617 new transiting planet candidates, identified in the Transiting
Exoplanet Survey Satellite (TESS) full-frame images (FFIs) observed during the Primary Mission (Sectors 1 -- 26). These candidates were initially detected by the Quick-Look Pipeline (QLP), which extracts FFI lightcurves for and searches all stars brighter than TESS magnitude $T = 13.5$ mag in each sector. However, QLP heavily relies on manual inspection for the identification of planet candidates, limiting vetting efforts to planet-hosting stars brighter than $T = 10.5$ mag and leaving millions of potential transit signals un-vetted. We describe an independent vetting pipeline applied to QLP transit search results, incorporating both automated vetting tests and manual inspection to identify promising planet candidates around these fainter stars. The new candidates discovered by this ongoing project will allow TESS to significantly improve the statistical power of demographics studies of giant, close-in exoplanets.
\end{abstract}


\section{Introduction}

NASA's Transiting Exoplanet Survey Satellite \citep[TESS;][]{Ricker2014} is the first nearly all-sky space-based transit search mission. Launched in April 2018, its Primary Mission observed $\sim$73$\%$ of the sky across 26 sectors, each lasting 27.4 days and covering a $24\degree\times96\degree$ field of view. Observations were taken with two data collection modes: 2-minute sampled ``postage stamps'' image cutouts centred on $\sim$20,000 pre-selected targets per sector, and 30-minute sampled full-frame images (FFIs) covering the entire TESS field of view. By the end of this two-year mission, TESS identified 2,241 exoplanet candidates \citep{Guerrero2021}, known as TESS Objects of Interest (TOIs).

A major pipeline for the search and analysis of planet candidates in TESS FFIs is the Quick-Look Pipeline \cite[QLP;][]{Huang2020, Kunimoto2021} at the TESS Science Office (TSO) at MIT. QLP performs multi-aperture photometry to extract lightcurves for all targets with TESS magnitude $T < 13.5$ mag from TESS FFIs. At the end of each sector, all available data for targets observed in that sector are stitched together, and a box-least squares (BLS) transit search \citep{Kovacs2002} implemented in \texttt{vartools} \citep{HartmanBakos2016} is run on these multi-sector lightcurves. Following BLS, QLP applies basic detection criteria to identify transit candidates: at least 5 points in transit, a signal-to-pink noise ratio of at least 9, and a signal-to-noise ratio (S/N) of at least 5 (if $T < 12$ mag) or 9 (otherwise). 

QLP then performs automated triage for passing signals around stars with $T < 10.5$ mag using AstroNet-Triage \citep[Moldovan et al., in prep;][]{Yu2019}, a neural network for distinguishing eclipsing or transiting objects from noise, intrinsic stellar variability, and contact binary stars. QLP operators manually review candidates passing AstroNet (typically a few hundred per sector), produce vetting reports, and deliver the reports to TOI vetters for further inspection. Candidates passing group vetting are alerted as new TESS Objects of Interest (TOIs) on the TOI Release Portal.\footnote{\url{https://tev.mit.edu/data/}}

The QLP team's choice to inspect signals only around stars brighter than $T = 10.5$ mag is primarily motivated by the fact that fainter searches would overwhelm operators and vetters with the number of candidates needing manual review. Indeed, over the entire Primary Mission, we found that 2,507,460 BLS signals passed QLP's detection criteria. After running AstroNet, we identified 686,242 transit candidates that would nominally have needed manual inspection had the vetting magnitude limit been $T = 13.5$ mag.

In this paper, we describe our application of an additional, automated vetting pipeline following triage by AstroNet for vetting these QLP threshold crossing events obtained on fainter stars. This stage was largely inspired by the Robovetter, a fully automated vetting tool first used to compile Kepler's DR24 catalog \citep{Coughlin2016} and again for DR25 \citep{Thompson2018}. We also describe the final stage of manual vetting which resulted in the identification of over a thousand new faint star TOIs.

\section{Identification of Planet Candidates}

\subsection{Automated Vetting}

Given that our vetting pipeline is not yet fully automated, our candidacy test thresholds were chosen empirically with the main purpose of reducing the number of signals needing manual inspection.

For any tests that required host star properties as an input, we adopted those properties from the TESS Input Catalog (TIC) v8.1 \citet{Stassun2019}. We assumed solar values when stellar properties were not available.

\subsubsection{Sine Wave Identification}

BLS searches are often confounded by stellar variability with strong quasiperiodic and sinusoidal components over short timescales. Some eclipsing binaries may also feature strong ellipsoidal variations in the lightcurves. To identify these cases, we fit a sine wave to the lightcurve with periods fixed to half, exactly, and twice the BLS period using the nonlinear least-squares fitting package \texttt{lmfit} \citep{Newville2016}, letting the amplitude and phase of the sine wave vary. We quantified the significance of the fit by dividing the fitted amplitude by the uncertainty in the amplitude, and rejected any candidates with more than 15$\sigma$ significance. 

\subsubsection{Model-Shift Uniqueness Test}

We employed the suite of model-shift uniqueness tests described in \citet{Coughlin2017} to do the bulk of automated vetting. In summary, Model-shift uses a transit model as a template to measure the amplitude of transit-like events at all phases in the phase-folded lightcurve. The procedure measures the significance and phases of the primary transit event, secondary and tertiary events, and the most significant positive flux (inverted transit-like) event. Model-shift also provides metrics such as the significance in the difference between odd and even transit events, and thresholds for determining if an event is significant compared to the noise level of the lightcurve. Overall, Model-shift allows for the identification of false positives due to noise and systematics (identified as signals that are non-unique compared to other events in the lightcurve, or with non-transit-like shapes), and those of astrophysical origin (identified as signals with significant secondary events or odd versus even depth differences).

Model-shift has been used in several automated vetting pipelines, including the Kepler Robovetter, the Discovery and Vetting of Exoplanets (DAVE) tool designed for K2 \citep{Kostov2019}, and the TESS-ExoClass (TEC) detection filter\footnote{\url{https://github.com/christopherburke/TESS-ExoClass}} used to vet TESS candidates in the TESS Science Processing Operations Center \cite[SPOC;][]{Jenkins2016} pipeline. To determine which candidates passed each test, we used the metric thresholds suggested in \citet{Coughlin2017}.

Model-shift requires a transit model fit to the lightcurve as an input. We fit a \citet{MandelAgol2002} quadratic limb-darkening transit model for each candidate\footnote{Adapted from \url{https://www.lpl.arizona.edu/~ianc/python/}}, parameterized by the orbital period ($P$), transit epoch ($T_{0}$), ratio of planet-to-star radii ($R_{p}/R_{s}$), distance between planet and star at midtransit in units of stellar radius ($a/R_{s}$), and impact parameter ($b$), with circular orbits assumed. Limb darkening parameters were taken from \citet{Claret2017} based on stellar parameters from the TICv8.1 stellar catalog \citep{Stassun2019}. To speed up the fit process, data more than two transit durations from the BLS-inferred center of each transit were ignored.

\subsubsection{Candidate Too Large}

An indicator of an eclipsing binary (EB) false positive is an extremely deep eclipse, which implies that the radius of the eclipsing object is too large to be planetary. We multiplied the $R_{p}/R_{s}$ fit results by the stellar radius of the target star, when available, and failed candidates with $R_{p} > 30 R_{\oplus}$.

\subsubsection{V-Shape Test}

While some EBs can be readily identified by their significant secondaries or large sizes, sufficiently grazing EBs ($b \gtrsim 1$) can have shallow eclipse depths that appear transit-like. These EBs can still be identified by both the depths and shapes of their eclipses. We adopted the Kepler DR25 Robovetter V-shape metric \citep{Thompson2018}, which required passing candidates have $R_{p}/R_{s} + b < 1.04$. We slightly relaxed this requirement to $R_{p}/R_{s} + b < 1.1$ to avoid inadvertently failing high-impact-parameter gas giants, especially those transiting small stars.

\subsubsection{Depth-Aperture Correlation}

QLP produces multiple lightcurves for each star, extracted via circular aperture photometry using different radii. Larger apertures are appropriate for brighter targets in order to capture most of their flux. However, larger apertures also include more flux from nearby stars. This flux contamination can result in false positives from nearby eclipsing binaries (NEBs), whose deep eclipses can appear transit-like in the target's lightcurve. An indication of NEB contamination is an increase in transit depth with aperture size.

We compared the lightcurves from Apertures 1 (radius of 2.5 pixels), 2 (3.0 pixels), and 3 (3.5 pixels) from QLP, where Aperture 1 is the default choice for the faintest stars ($T < 11.5$ mag) and Aperture 2 is the default choice for most brighter stars ($11.5 < T < 8.5$ mag). We measured the transit depths as the mean of the central 30 minutes of each phase-folded lightcurve, and estimated the noise by 1.4826 times the median absolute deviation (MAD) of the out-of-transit lightcurve. 1.4826 is a conversion factor to put the variability on the same scale as a Gaussian standard deviation. We failed the candidate if an increase in aperture size resulted in a more than $1\sigma$ increase in transit depth. While this is a strict threshold, this was the only flux-level test capable of removing off-target signals, and we found that even small aperture-depth differences were reliable indicators of centroid offsets.

\subsubsection{Centroid Offsets}

The depth--aperture correlation test can identify NEBs well-separated from the target, but pixel-level analysis is needed for closer contaminants. A powerful method for the identification of NEBs is the difference image technique described in \citet{Bryson2013}. In summary, the average of in- and out-of-transit pixels surrounding the target are found from a candidate's transit ephemerides and duration. The out-of-transit image represents a direct image of the field surrounding the target star. Meanwhile, the difference between the in- and out-of-transit images should appear star-like at the location of the transit source, assuming the transit is the explanation for any difference in flux. If the field is relatively uncrowded and the target star is indeed the source of the transit signal, the direct and difference images should appear similar.

We generated difference images using \texttt{TESS-plots}\footnote{\url{https://github.com/mkunimoto/TESS-plots}}, which includes a tool for difference image generation for candidates in TESS FFIs following the technique outlined in \citet{Bryson2013}. Due to the computational expense of making and storing these data products, we only produced difference images for the 37,022 signals passing all previous tests. We also employed a simple, fast method to locate the source of the transit as the flux-weighted centroid of the difference image, calculated using \texttt{center\textunderscore of\textunderscore mass} from \texttt{scipy.ndimage}. We failed candidates for which the centroid was more than 1 pixel offset from the pixel location of the target star as predicted by the \texttt{tess-point}\footnote{\url{https://github.com/christopherburke/tess-point}} high precision pointing tool.

\subsection{Manual Vetting}

\begin{figure*}[t!]
    \centering
    \includegraphics[trim=3cm 1cm 3cm 1cm, clip, width=\textwidth]{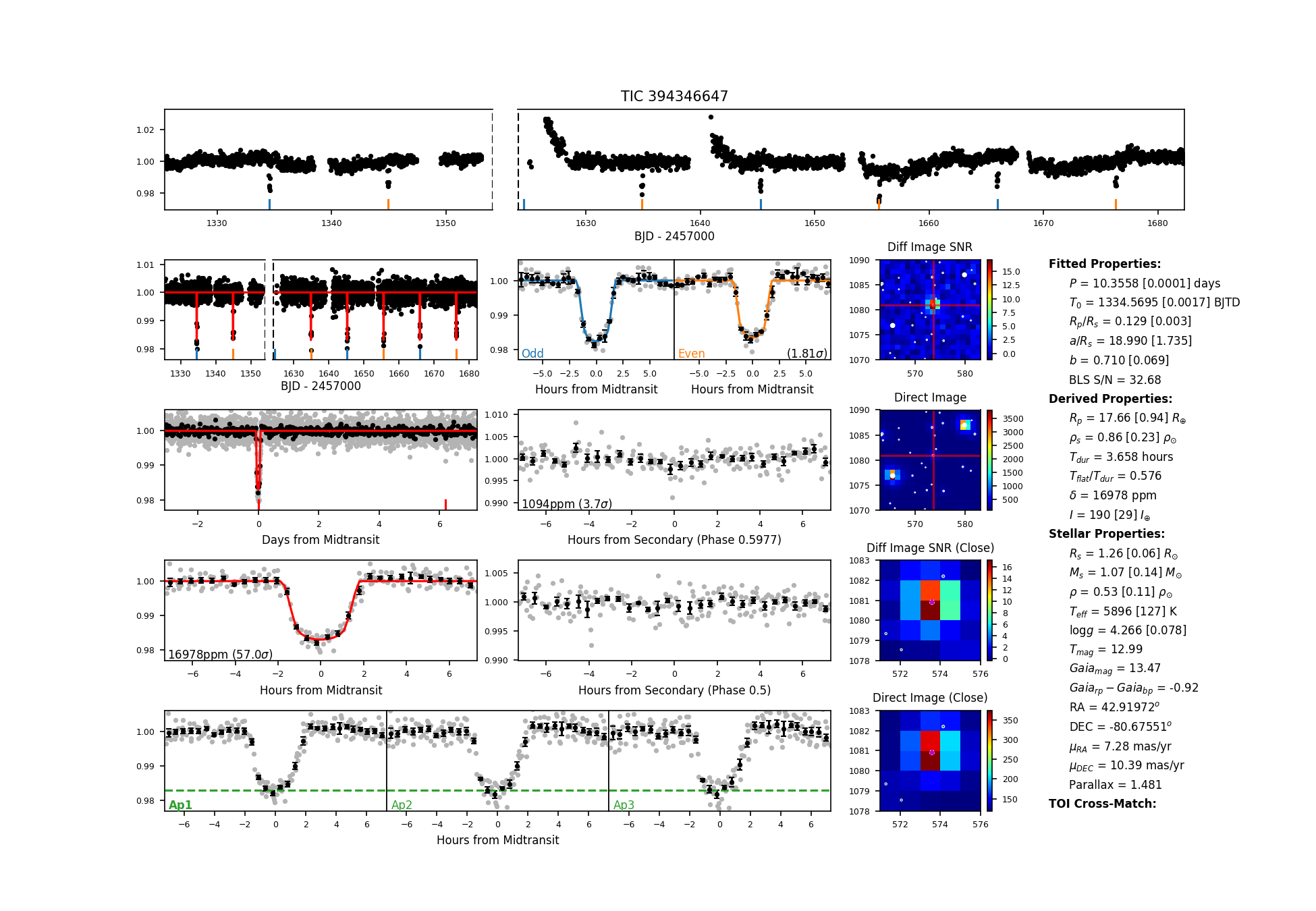}
    \caption{The vetting page for the $10.4$-day planet candidate orbiting TIC-394346647 (now TOI-2620), containing data across Sectors 1, 12, and 13. \textbf{Top row:} raw lightcurve, with odd and even transits marked in blue and orange, respectively. \textbf{Second row, first panel:} detrended lightcurve with the transit model in red. \textbf{Second row, second panel:} odd and even transit phase diagrams. \textbf{Third row, first panel:} full phase diagram. \textbf{Third row, second panel:} phase diagram zoomed into the most significant secondary event from Model-shift. \textbf{Fourth row, first panel:} phase diagram close-up to transit event with the transit model in red. \textbf{Fourth row, second panel:} phase diagram close-up to phase 0.5. \textbf{Fifth row:} phase diagram close-up to transit event in three different apertures. \textbf{Second last column:} Full $20\times20$ pixel view of difference and direct images for Sector 13, followed by a close-up of the central $5\times5$ pixels. The target star is marked by a pink star. Nearby stars down to 4 magnitudes fainter are plotted as with circles. \textbf{Last column:} fitted and derived parameters from the transit model, BLS, and TIC, and the stellar catalog, with an indication of any matches to known TOIs.}
    \label{fig:394346647}
\end{figure*}

6,936 candidates passed the automated vetting pipeline and moved to manual vetting. We produced vetting report pages to aid in the visual inspection of each passing candidate. An example is shown for the 10.4-day planet candidate orbiting TIC-394346647 (now TOI-2620) in Figure \ref{fig:394346647}, which is a $T = 12.99$ mag star observed in the Primary Mission during Sectors 1, 12, and 13. The vetting report includes plots of the full raw and detrended lightcurves, phase diagrams centered on the transit, odd versus even transits, the most significant secondary, and the transit in three different apertures. The report also includes the direct and difference images for by-eye identification of centroid offsets, and a list of relevant transit, planet, and stellar properties.

Two vetters independently reviewed each report that did not correspond to an already known TOI, and assigned either P (planet), F (false positive), or U (undecided) labels. Disagreements between vetters were discussed and resolved in a group vetting session. For signals observed in multiple sectors, we defaulted to the disposition from its latest observed sector.

Common candidates that needed to be resolved through group discussion were those in crowded fields. One issue with high levels of crowding is an increased probability of NEB contamination, although manual inspection of the depth--aperture correlation and difference image plots were useful in identifying these cases. However, crowding can also cause transit depth dilution, meaning our planet radii would be under-estimated. Because many of the candidates we reviewed were giants, for which modest levels of dilution could increase their radii to non-planetary values, we tended to treat candidates critically in crowded fields.

A QLP operator produced standard QLP reports for all signals labeled with P, with the inclusion of any available TESS Extended Mission data for targets that had been re-observed since the Primary Mission. Each QLP report was then reviewed by the operator as a final manual check that the signal was still consistent with a planet interpretation before delivering the candidates to the TOI release portal, from which they were alerted as a TOI.

Nine of the Faint Star TOI hosts had been observed with two-minute cadence observations during the TESS Primary Mission.\footnote{\url{https://tess.mit.edu/observations/target-lists/}} We checked each of these targets for SPOC Threshold Crossing Events (TCEs)\footnote{\url{https://archive.stsci.edu/tess/bulk_downloads/bulk_downloads_tce.html}}, finding that six of the nine stars did not result in TCEs. All of these candidates had low S/N or only 1 -- 2 transits per sector. We believe that these candidates went undetected by the SPOC pipeline primarily because fewer sectors of 2-min data were available for SPOC than sectors of FFI data were available for QLP.

Two of the remaining three hosts had SPOC TCEs which failed the Data Validation (DV) stage of the SPOC pipeline, though the TCEs did not match our planet candidates. TIC-229786610 gave a non-transit-like $P = 0.27$-day TCE signal in SPOC multi-sector, whereas QLP revealed a $P = 22.2$-day, S/N = 10 candidate later alerted as TOI-4113.01. TIC-169461816 also gave a non-transit-like $P = 0.27$-day SPOC TCE in both single- and multi-sector searches, whereas QLP detected a single transit at $T_{0} = 2458729.2$ days (BJD). This candidate was later alerted as TOI-3563.01, and is an independent TESS detection of the known planet Kepler-448 c ($P = 17.9$ days, $T_{0} = 2454979.6$ days).

Finally, one Faint Star TOI matched a SPOC detection that passed DV. TIC-301160638 (TOI-3487) revealed the same two-transit, $P = 16.0$-day signal in both SPOC and QLP data. Vetters likely failed the signal from the SPOC report due to a deep event $\sim$6 days after the first transit, which could indicate that the transits of TOI-3487.01 are secondaries of an eccentric eclipsing binary star. However, this event landed in data flagged as poor quality by QLP, and the out-of-transit flux surrounding the event in SPOC data is not flat. There were also not enough TESS observations to confirm that the event repeated. Because all other indications from both SPOC and QLP showed that TOI-3487.01 was consistent with a high S/N, on-target planet candidate, we opted to alert the signal.

\section{Results and Discussion}

A total of 1,617 TOIs were alerted from the QLP Primary Mission faint star search. These TOIs are tracked in the TOI Catalog\footnote{\url{https://tev.mit.edu/data/}} by the comment ``found in faint-star QLP search.''

\subsection{Comparisons with Other TOIs}

\begin{figure}[h!]
    \centering
    \includegraphics[width=\linewidth]{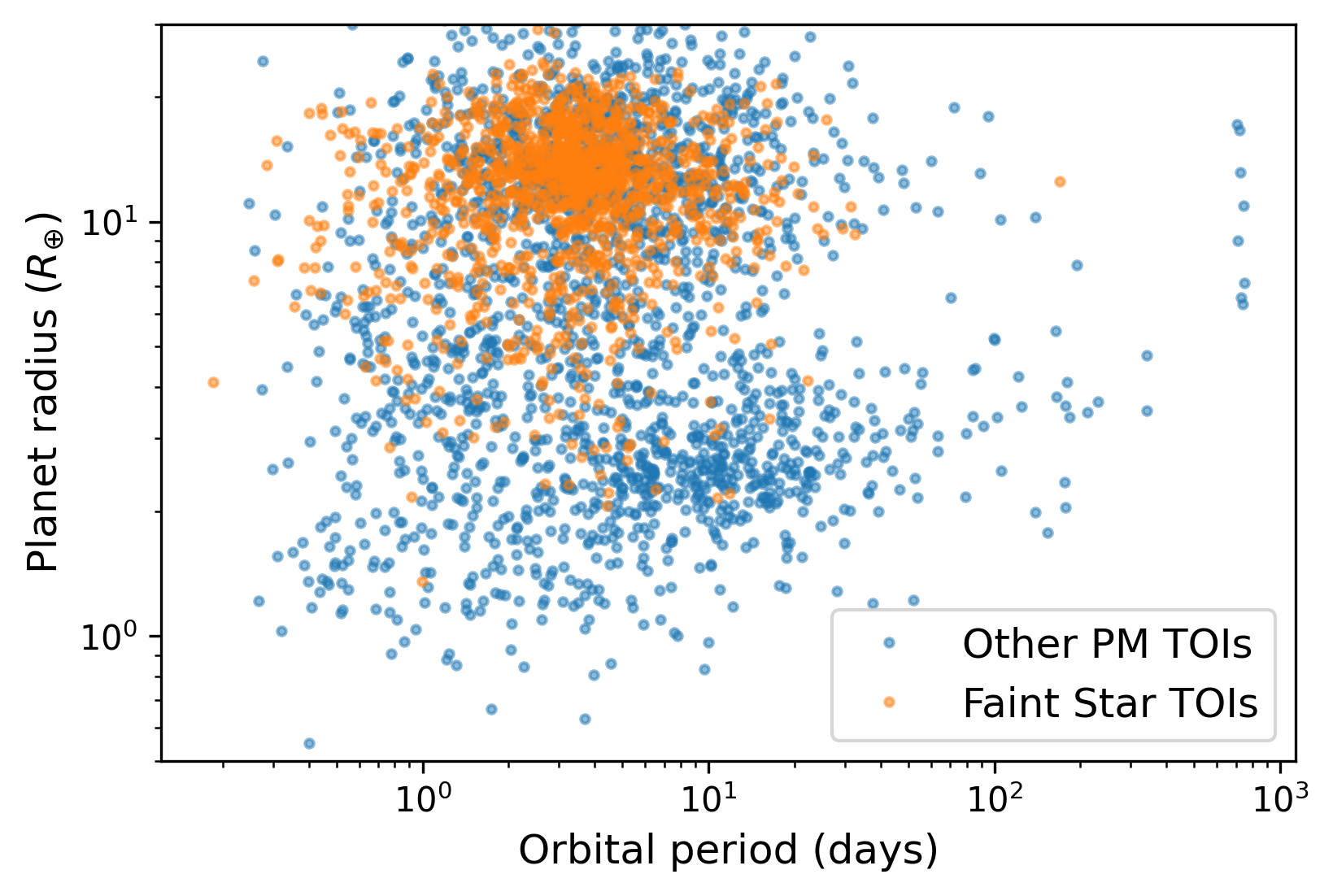}
    \includegraphics[width=\linewidth]{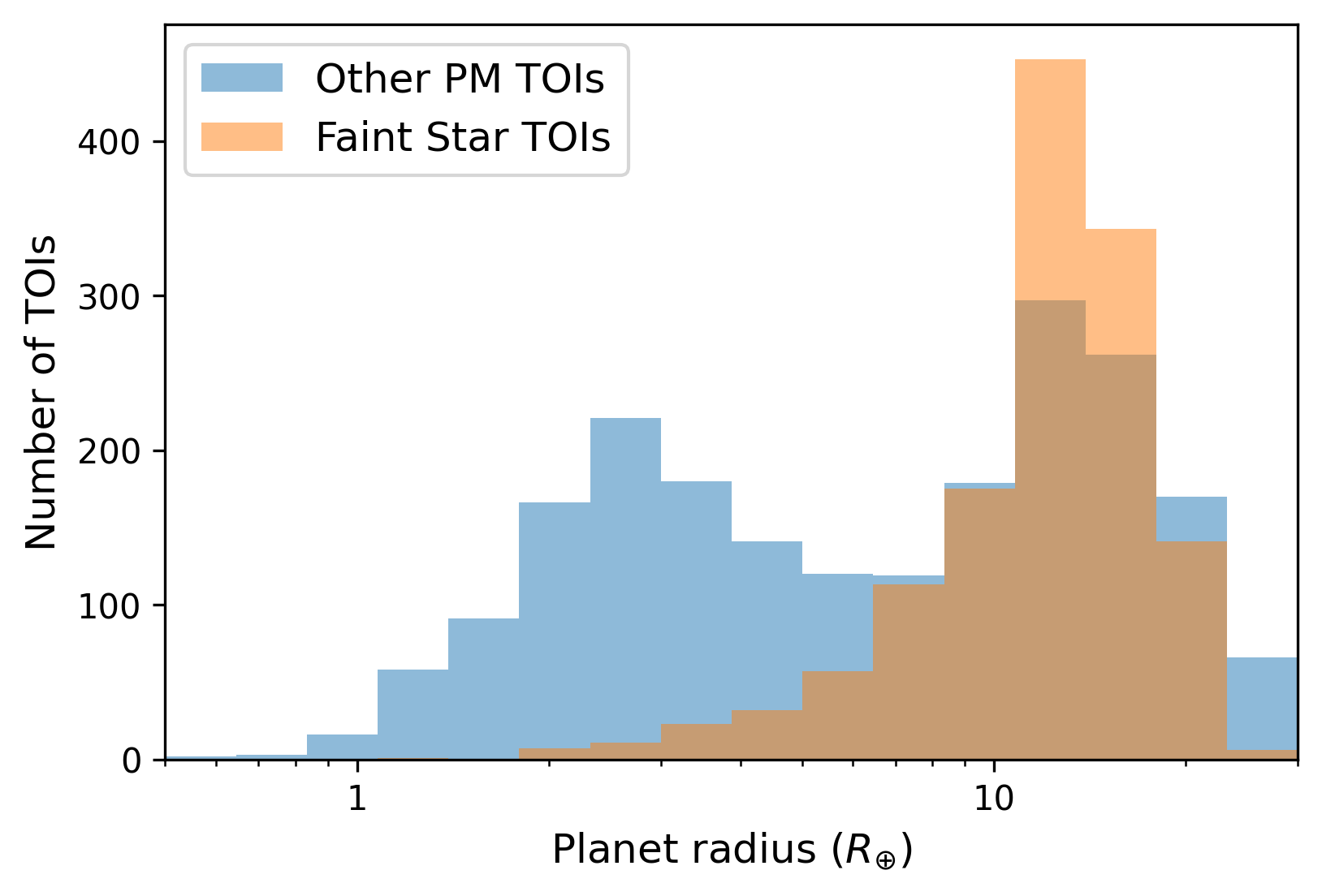}
    \includegraphics[width=\linewidth]{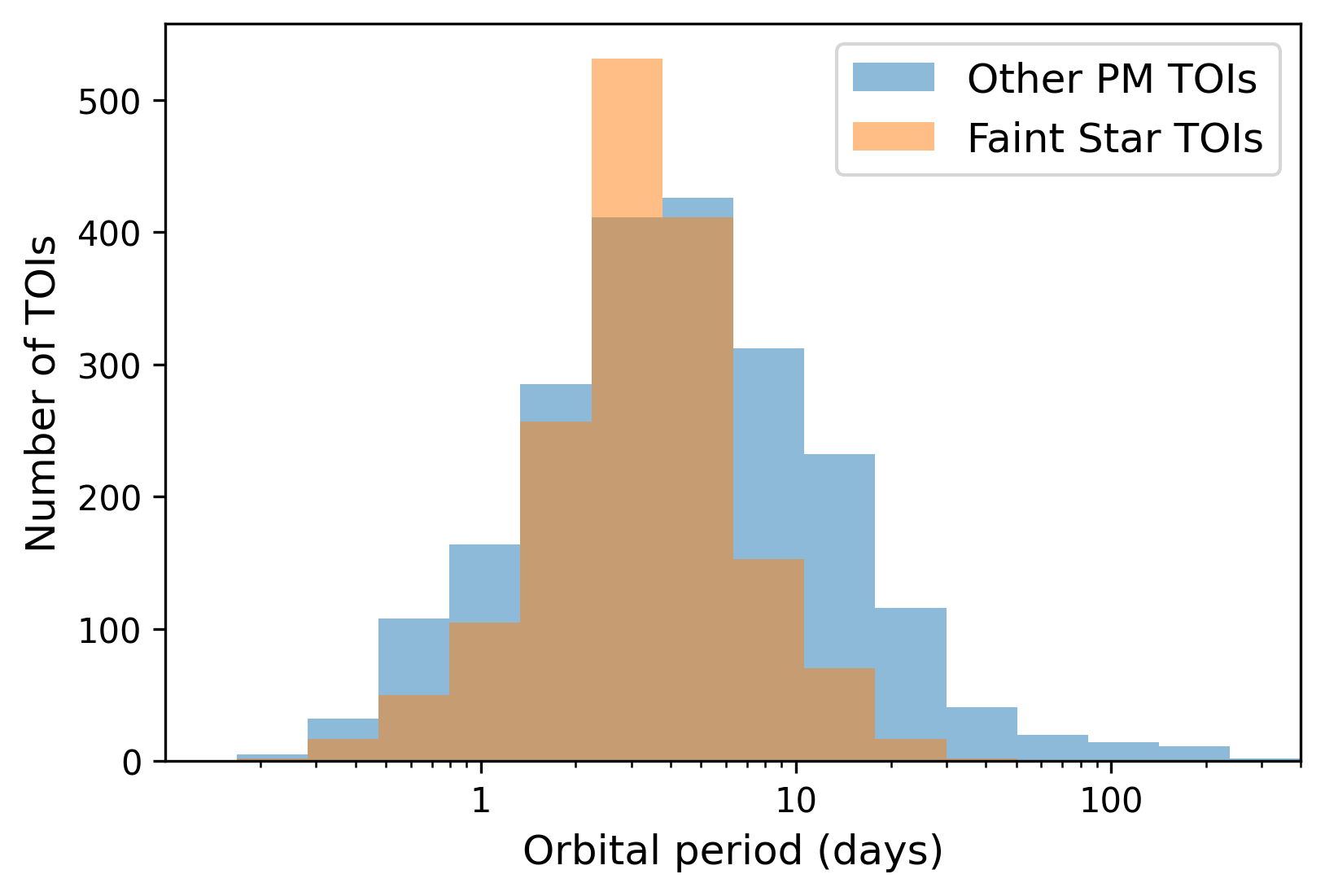}
    \caption{A comparison of TOI properties from the TOI catalog, with the Faint Star TOIs in orange and other TOIs from the Primary Mission \citep{Guerrero2021} in blue. \textbf{Top:} TOIs plotted in period-radius space. TOIs larger than $30 R_{\oplus}$ are not shown as they are highly unlikely to be planetary. \textbf{Middle:} histograms of planet radii. \textbf{Bottom:} histograms of orbital periods. The faint star TOIs are predominantly giant candidates ($R_{p} > 10 R_{\oplus}$) and those with short orbital periods ($P < 10$ days).}
    \label{fig:per_rad}
\end{figure}

Figure \ref{fig:per_rad} shows how the Faint Star TOIs compare with other Primary Mission TOIs across period and radius space, with properties adopted from the TOI Catalog\footnote{Accessed 2021 November 30 from \url{https://exofop.ipac.caltech.edu/tess/}}. The Faint Star TOIs are clearly dominated by giant planet candidates and those with short orbital periods. On average, the Faint Star TOIs have orbital periods 21\% shorter than other TOIs and 24\% larger transit duty cycles. The giant, close-in planet bias has two explanations: First, lightcurve precision worsens as one moves to fainter stars, with noise on transit timescales on the order of $1000$ ppm at $T \approx 13.5$ mag \citep{Huang2020}. Planet transits must therefore be very deep (several thousand ppm) to be detectable around these faint stars, which naturally favours the detection of large planets. We found that the transits of Faint Star TOIs are approximately 3.5 times deeper on average compared to the rest of the TOI process in order to compensate for the drop in the photometric precision caused by the increase in the limiting magnitude. Furthermore, luminous and therefore large stars are overrepresented in a magnitude-limited search, which also favours the detection of large planets. Second, we reviewed QLP results from the Primary Mission, for which most stars were observed in only a single 27.4-day sector. Given that multiple transits are needed to confirm periodicity, this baseline limits the majority of detectable signals to those with orbital periods less than half the length of a sector.

\begin{figure}[h!]
    \centering
    \includegraphics[width=\linewidth]{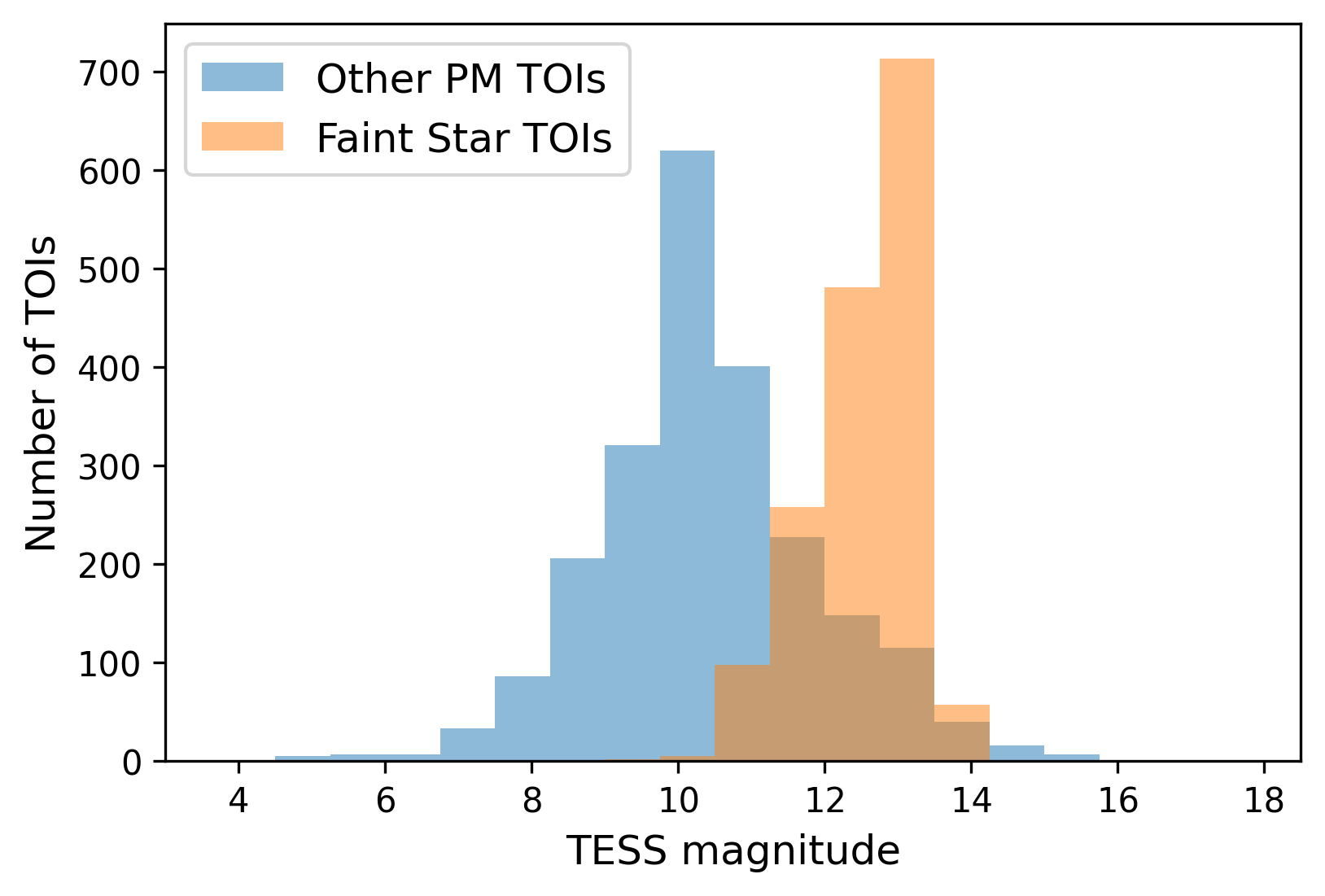}
    \caption{A comparison of the TESS magnitudes of TOI hosts, with the Faint Star TOIs in orange and other TOIs from the Primary Mission in blue. On average, Faint Star TOIs orbit stars 2.4 magnitudes fainter than other Primary Mission TOIs.}
    \label{fig:tmag}
\end{figure}

Compared to the rest of the Primary Mission TOI process, Faint Star TOIs are fainter on average by 2.4 magnitudes in the TESS band (Figure \ref{fig:tmag}). Consequently, with an average distance to the Solar system of 615 pc, Faint Star TOI hosts are three times more distant than rest of the TOIs. They are also preferentially hosted by large (Sun-like and larger) stars, with larger radius by 10\% on average. We expect that the absence of M-dwarfs will be addressed in our analysis of Extended Mission data (Daylan et al. in prep) thanks to the change of the FFI cadence from 30 to 10 minutes, given the relatively small size of M-dwarfs and correspondingly smaller transit durations of planets orbiting them.

\subsection{Scientific Implications}

Of the 1,617 Faint Star TOIs, 1,014 (63$\%$) can be considered hot Jupiter candidates ($R_{p} > 9 R_{\oplus}$, $P < 10$ days), nearly a factor of 10 more than the number of hot Jupiters found by the Kepler mission.\footnote{Based on the number of Confirmed and Candidate Kepler Objects of Interest listed on the NASA Exoplanet Archive, accessed 2021 September 4.} While some of these TOIs will unavoidably be false positives \citep{Santerne2013}, these findings demonstrate TESS's potential for significantly improving our understanding of hot Jupiter demographics. \citet{Yee2021} predicted that a faint magnitude-limited survey with TESS could increase the number of hot Jupiters over Kepler by an order of magnitude, and our Faint Star search supports this prediction. Because the identification of planet candidates by our pipeline still relies on biased manual inspection, we caution that the hot Jupiters discovered by the Faint Star (or regular TOI) process should not yet be used for statistical analysis. However, identification of TOIs early on gives the follow-up community time to statistically validate planets and identify false positives.

New TOIs are also promising targets for multi-planet searches, especially when TESS re-observes many of these stars in the current Extended Mission and beyond. The Kepler mission has shown that multi-planet systems are common \citep[e.g.][]{Yang2020}, and the orbital planes of additional planets in known transiting planet systems are highly likely to be aligned with our line of sight. While rare, companions to hot Jupiters are also valuable probes of hot Jupiter formation. Our faint star search has already uncovered a new TESS multi-planet system, TIC-352682207 (TOI-4010), a K dwarf which potentially hosts three close-in planets (Kunimoto et al. in prep).

\subsection{Promising Targets for Spectroscopic Follow-up}

One of the primary goals of the TESS mission is to discover small planets ($R_{p} < 4 R_{\oplus}$) suitable for mass measurements via ground-based radial velocity (RV) observations \citep{Ricker2014}. To prioritize targets for RV follow-up, we estimate the predicted RV semi-amplitude $K$ for each small planet using 

\begin{equation}
    K = 28.4 \bigg(\frac{M_{p}}{M_{J}}\bigg)\bigg(\frac{M_{p} + M_{s}}{M_{\odot}}\bigg)^{-1/2}\bigg(\frac{a}{\text{1 AU}}\bigg)^{-1/2}\text{ m s$^{-1}$}
\end{equation}

\noindent where $M_{J}$ is a Jupiter mass, $M_{p}$ is planet mass converted from $R_{p}$ using the mass--radius relations from \citet{ChenKipping2017}, $M_{s}$ is the stellar mass, and circular, edge-on orbits have been assumed.

Due to the nature of this work, the majority of our candidates are giants, or orbit stars too faint for RV follow-up. However, we find 10 planet candidates with $R_{p} < 4 R_{\oplus}$ and expected $K > 1$ m/s that orbit stars brighter than $V = 12$ mag. Of these, two (TOI-4110.01 and 4219.01) have $K > 5$ m/s. We also find three small planet candidates (TOI-2486.01, 2768.01, and 4010.03) with $K > 5$ m/s that orbit likely K and M dwarf stars brighter than $T = 12$ mag, which may be more accessible with spectrographs that operate in the red end of the spectrum \citep[e.g. MAROON-X;][]{Seifahrt2018}. For those interested in follow-up of larger planets, 99 of our planet candidates have $K > 10$ m/s and orbit stars brighter than $V = 12$ mag.

A secondary goal of TESS is to discover exoplanet targets amenable to atmospheric characterization. We adopted the framework developed by \citet{Kempton2018} to identify the most promising targets from our TOI list. \citet{Kempton2018} introduced the Transmission Spectroscopy Metric (TSM) and Emission Spectroscopy Metric (ESM), which quantify the expected S/N in transmission and thermal emission spectroscopy, respectively. Planet candidates satisfying $R_{p} < 4 R_{\oplus}$ and TSM $> 10$ or $R_{p} < 10 R_{\oplus}$ and TSM $> 90$ can be considered high-quality targets for transmission spectroscopy, while small planets with $R_{p} < 1.5 R_{\oplus}$ and ESM $> 7.5$ are good targets for emission spectroscopy \citep{Kempton2018}. We calculated the TSM and ESM for each of our planet candidates with $R_{p} < 10 R_{\oplus}$, and found 31 with TSM $> 90$.

One highlighted Faint Star TOI is the small planet candidate TOI-2486.01 ($R_{p} = 3.7 R_{\oplus}$, $P = 1.5$ days). TOI-2486.01 orbits a late K dwarf ($M_{s} = 0.66 M_{\odot}$, $R_{s} = 0.77 R_{\odot}$, $T_{\text{eff}} = 4205$ K) which is one of our brighter targets in the TESS band ($T = 11.1$ mag). With both $K = 9.8$ m/s and TSM = 92.4, we consider this our best small candidate for spectroscopic follow-up. TOI-2486.01 was also independently detected by \citet{Montalto2020}.

\section{Concluding Remarks}

We have described the identification of 1,617 new TOIs hosted by stars brighter than $T = 13.5$ mag using FFIs collected in the TESS Primary Mission. These planet candidates were initially detected by the QLP transit search \citep{Huang2020}, but were not inspected because QLP vetting is performed only for targets brighter than $T = 10.5$ mag. Our vetting included an independent and automated vetting pipeline, followed by manual review to determine the final list of TOIs to be alerted. Overall, the TOIs resulting from this work constitute roughly 40\% of all TOIs from the TESS Primary Mission. With the identification of more than 1000 new hot Jupiter candidates, these Faint Star TOIs will allow TESS to significantly improve understanding of giant, close-in exoplanets.

The vetting pipeline described in this work enabled the inspection of $\sim$700,000 signals which would otherwise have to be manually reviewed by QLP operators. We plan to extend this pipeline further into a fully automated planet vetting pipeline designed specifically for TESS by tuning our metric thresholds and adding new tests in the future. Aside from further reducing manual workload on vetting planet candidates, this will also apply to exoplanet occurrence rate studies, which require uniformly produced planet catalogs and accurately characterized selection functions.

Our review of lightcurves of faint stars from the TESS Extended Mission (Cycles 3 and 4) is ongoing and will be published in a future work.

\begin{acknowledgments}
This paper utilizes data from the Quick Look Pipeline (QLP) at the TESS Science Office (TSO) at MIT. The TESS mission is funded by NASA’s Science Mission Directorate.

This research has made use of the Exoplanet Follow-up Observation Program website, which is operated by the California Institute of Technology, under contract with the National Aeronautics and Space Administration under the Exoplanet Exploration Program. 
\end{acknowledgments}

\software{\texttt{lmfit}~\citep{Newville2016}, \texttt{matplotlib}~\citep{Hunter2007}, \texttt{numpy} \citep{Harris2020}, \texttt{scipy} \citep{Virtanen2020}} 

\bibliography{refs}

\end{document}